# Synergistic Development of Cybersecurity and Functional Safety for Smart Electric Vehicles

## Siddhesh Pimpale

**Abstract:** The introduction of Smart Electric Vehicles (SEVs) represents an increasingly disruption on automotive area, once integrates advanced computer and communication technologies to highly electrical cars, which come with high performances, environment friendly and user friendly characteristics . But the increasing complexity of SEVs prompted by greater dependence on interconnected systems, autonomous capabilities and electrification, presents new challenges in cybersecurity as well as functional safety. The safety and reliability of such vehicles is paramount, as unsafe or unreliable operation in either case represents an unacceptable risk in terms of the performance of the vehicle and safety of the passenger. This paper investigates the integrated development of cybersecurity and functional safety for SEVs, emphasizing the requirement for the parallel development of these domains as components that are not treated separately. In SEVs, cybersecurity is quite crucial in order to prevent the threats of hacking, data breaches and unauthorized access to vehicle systems. Functional safety ensures that important vehicle functions (braking, steering, battery control, etc.) keep working even if some part fails. This convergence of functional safety issues with cybersecurity issues is becoming more crucial, since a security incident can result in a failure of catastrophic consequences for a functional safety system and, conversely. This paper reports the current state of cybersecurity and functional safety standards for SEVs, highlighting challenges that include the weaknesses of communication networks, the potential security threats of over-the-air updates, and the demand for real-time responsive systems for failure. In this paper we suggest a unified framework that combines cybersecurity measures with functional safety principles, with an aim at providing secure systems in which no danger can arise from the attack while maintaining the primary safety level in SEVs. The approach underlines a multi-layered defense with monitoring, secure authentication as well as a thorough test process addressing safety and security. This article uses case studies and industry practices to illustrate the advantages of taking an integrated approach to cybersecurity and functional safety, which can result in SEVs that are more robust against cyber-attacks and functional failures. Finally, recommendations for future research and development are provided, highlighting the urgent requirement for collaboration between carmakers, regulators, and cybersecurity professionals to guarantee SEVs are some of the safest, secure, and reliable vehicles for the fast-paced automotive industry.

## Introduction

The commercial rollout of Smart Electric Vehicles (SEVs) in the automotive domain exhibits a major step-change brought about by electrification, automation and connectivity. SEV not only holds the potential to reduce emissions and improve energy usage, but will also enable fully autonomous and connected mobility solutions. However, the growing sophistication of SEVs, thanks to the incorporation of sophisticated systems such as driver-assist functionalities, over-the-air updates, vehicle-to-

*spimpale848@gmail.com*

*Dana Inc, USA*

everything (V2X) communication, and autonomous functionalities and features, has posed new vulnerabilities, particularly in the areas of cybersecurity and functional safety (Sharma & Singh, 2022).

If the emphasis on functional safety in SEVs used to be largely directed towards managing the safety of critical vehicle functions like braking, steering, or energy management, the emergence of connected and autonomous vehicles (CAVs) points to increased attention to cybersecurity. Vehicles are protected from unauthorized access, hacking as well as malicious attacks that may interfere with the vehicle operation or pose threats to passengers safety by the cybersecurity (Xia et al., 2022). Such risks are amplified with the rise of SEVs linked to external world such as the Internet, smart grids, as well as other SEVs, which exposes them to different cyber-attacks (Fang et al., 2021).

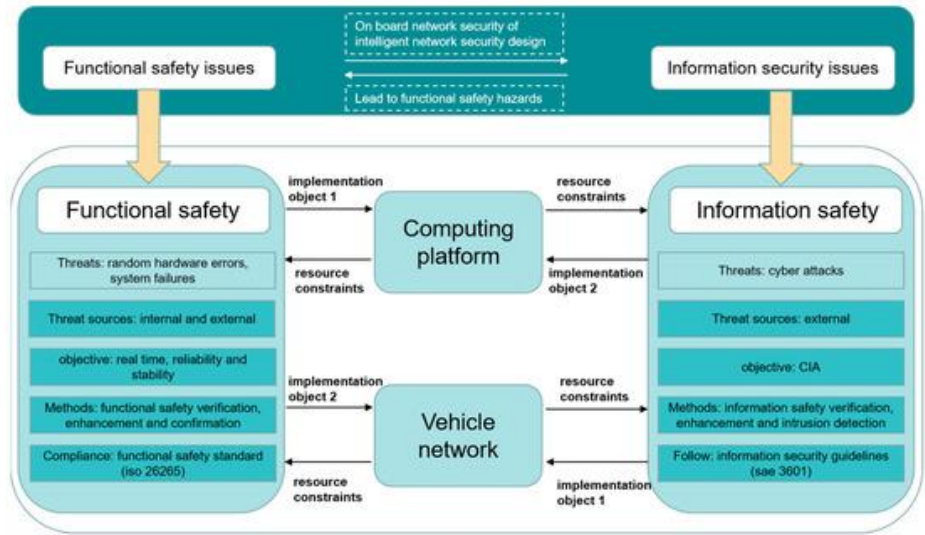

**Figure 1.** Relationship between function security and information security in intelligent network-connected vehicles.

Performance safety is the capability of the vehicle to stay safely operational after a failure of part of its systems and is guided by international standards such as ISO 26262. The above standard details the requirements which must be followed to develop, verify and to integrate safety-related systems for automotive applications. These standards are of even more importance in SEVs as the failure of critical components in SEVs is likely to have catastrophic outcome such as accidents or malfunctioning of the vehicle systems (Chaudhary et al., 2020). Similarly, cybersecurity norms (including ISO/SAE 21434) focus on determining, handling and reducing risks regarding cybersecurity in automobile systems. These standards serve as a guide to ensure that automotive systems are resistant to cyberattacks and focus on maintaining vehicle security throughout the life cycle (Marupaka et al., 2021).

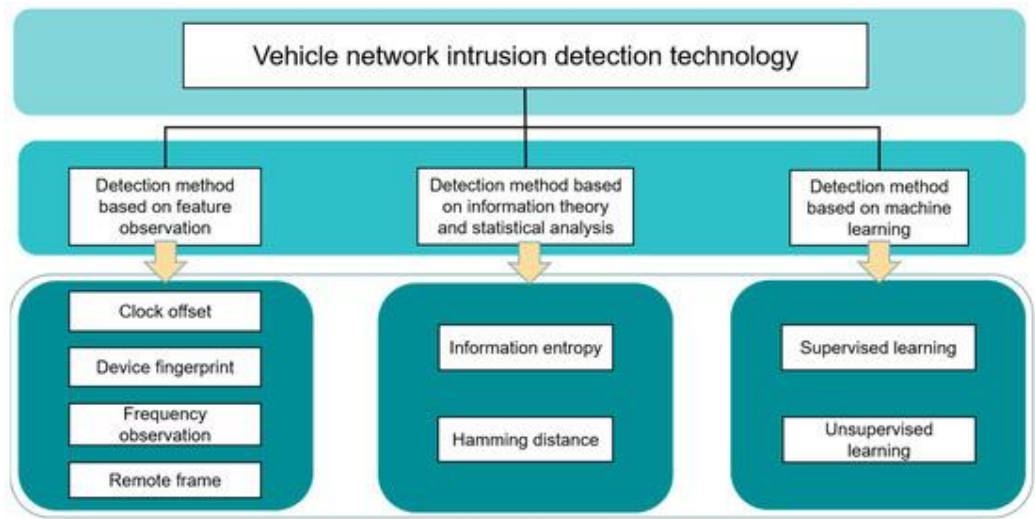

**Figure 2. Intrusion detection technology classification of vehicle network.**

While the relevance of both domains is evident, there are usually separate discussions on cybersecurity and functional safety in SEVs. This dual goal is necessary, but presents some challenges to maintain on one hand a level of cybersecurity, which do not endanger the vehicle safety, and make sure on the other one that the safety devices do not expose the system to attack. For example, an effort to strengthen cybersecurity by implementing active software updates from a distant location can inadvertently expose mission- and safety-critical systems to cybersecurity risk if not rigorously managed and verified during the software update lifecycle. Likewise, safety critical systems that exploited data inputs of connected origin may be exposed to cybersecurity-related risks, when these risks are not adequately met (Liu et al., 2022).

It is essential that these two vital areas converge in order to result in an all-encompassing, dependable and secure SEV construction. The focus of this paper is to study the synergistic development of cybersecurity and functional safety in SEVs, and particularly how the robustness of each one is integrated together to form a unified framework that guarantees security as well as safety of the system. The model presented in this paper propose some solutions to the open issues and suggests integrated, end-to-end security measures which shin protect the vehicle both inside (components, physical layers) and toward the outside world (external software systems, external communications systems) and on which the vehicle relies.

To mitigate these challenges, we review and analyze the status quo of cybersecurity and functional safety in SEVs, discuss existing industry practices, and provide a unified framework for integrating cybersecurity and functional safety standards for SEVs. This framework should help to inform the design, testing and operation of SEVs that are both safer and more secure; and, specifically, to address risk management, monitoring in real-time, and multi-layered protective measures that can serve to guarantee the safe deployment and reliable operations of these complex systems.

## Literature Review

Smart Electric Vehicles (SEVs) are on the horizon, reshaping the automobile industry with technological breakthroughs in electric propulsion, autonomous driving systems, and vehicle-to-everything (V2X) connectivity. Keeping SEVs safe, robust and secure also becomes more complicated as these systems grow. These are of particular importance in these vehicles which rely extensively on sophisticated electronic systems and communication systems. This review will focus on the convergence of cybersecurity and functional safety in SEVs, trends and challenges and how these two hot areas can be combined to make vehicle systems safer and trusted.

### 1. Cybersecurity in Smart Electric Vehicles

The proliferation of connected and autonomous technologies in SEVs has made SEVs increasingly susceptible to cyber-attacks that threaten the safety and privacy of the vehicle occupants. SEVs have multiple connectivity systems, such as V2V communication and V2I communication, OTA update and cloud aboard data storage, which enlarge the attack surface resulted from cyber risks (Fang et al., 2021). Such threats can be unauthorized entry into essential system of the vehicle, for example, control units, navigation or power management, to perform malicious operations, e.g., remote control, data extraction or system failure.

Automotive security is mainly concerned with preventing the unauthorized access to the vehicle's communication networks while also guaranteeing the integrity of the data and controls exchanged between the vehicle and the outside world. Multiple cybersecurity standards are established that aim to mitigate these risks, most prominently ISO/SAE 21434 that offers recommendations to manage cybersecurity across the vehicle lifecycle (Marupaka et al., 2021). This specification aims to assess and address cybersecurity vulnerabilities of intelligent transportation systems, with a special concern on secure communication protocols, strong identity, and ongoing threat detection.

Furthermore, Fang et al. (2021) point out that in SEVs, cybersecurity is not only about protection against attacks, but also about the resilience of the vehicle systems to cyber-triggered failures. For instance, communication security between vehicle internal components and other networks on the road is necessary so that adversaries cannot exploit communication links of the system in a manner that leads to hazardous situations. Robustness of these networks is extremely important as any failure can influence those functions of cars which affect safety.

## 2. Functional Safety in Smart Electric Vehicles

Regarding of functional safety in SEVs, it considers systems design and construction that necessary for certification of vehicle functions for braking, steering and energy management ensuring that they are still safely, even in case of faults or failures. The automotive industry uses global safety standards like ISO 26262, which defines requirements for the functional safety of electrical and electronic systems in automobiles (Chaudhary et al., 2020). These standards specify procedures in terms of what is required to assure that safety-related vehicle systems are designed, tested, and proven safe to prevent and mitigate hazards due to failures in these systems.

For SEVs, functional safety becomes of highest relevance since we are dealing with more complexity in the electrical powertrain and autonomous driving functionalities. The automotive industry has consequently established rigorous procedures for designing fail-safe systems, and in the operational context these systems commonly are subjected to extensive testing to ensure that they remain safe even if they fail. For SEVs, the battery management system (BMS) is also an essential component for functional safety as it observes the state of charge, temperature, and health of the battery. One potential risk is a malfunction of the battery management system, which may cause overcharging, overheating, and in the worst case, possible runaway.

## 3. Security and Safety Integration

The automotive domain is digitalizing and becoming connected, which makes it next to mandatory to integrate cyber security and functional safety. Although the two domains evolved separately, they must merge in SEVs, as the failure of one system can put the other at risk.

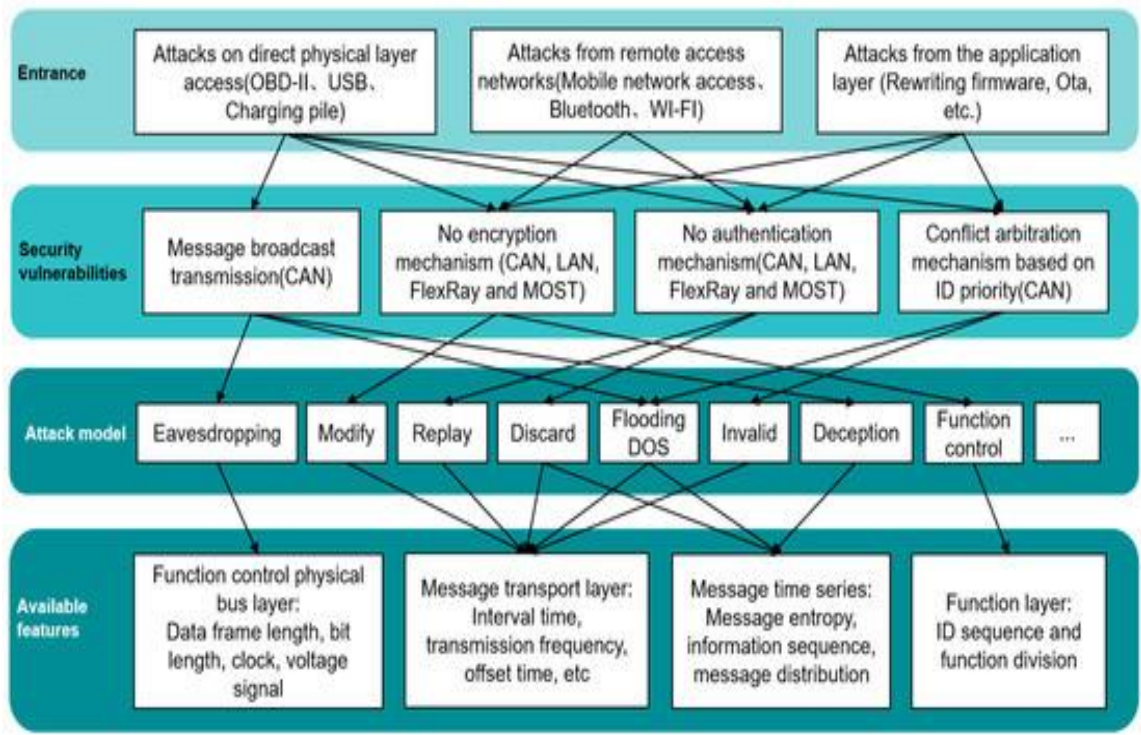

**Figure 3. Relationship between attack entry, security vulnerability, and exploitable features of intrusion detection in vehicle networks.**

## 4. Challenges and Future Directions

Cybersecurity and functional safety integration in SEVs are identified with various challenges. The systems themselves are also too difficult to be implemented. However, as SEVs add in more autonomous functionalities and networking capabilities, the number of devices linked together becomes very challenging to handle and secure (Sharma & Singh, 2022). Furthermore, any attempt to address apprehensions in a dynamic environment come with the "requisite real-time monitoring and threat detection [that] are the necessary components to ensure the continued safety and security." As machine learning and AI continue to develop, tomorrow's systems could use advanced predictive analytics to identify potential threats as they start growing, rather than when they reach critical status.

The changing regulatory environment is adding to the difficulty here. With technology of cars being updated, facilitating the control centers with current data corresponding to the quality and quantity of the ecological / meteorological parameters seems making beneficial weak point as (Sharma & Singh, 2022) mentioned. This calls for agility in standards creation in the face of dynamic SEV technology. A joint effort will be key in defining and implementing successful practices that combined will guarantee safety and security, whit standards driving such an agreement between carmakers, regulatory agencies and security researchers.

The literature shows the vital importance of the fusion between Cybersecurity and Functional Safety for both design and operation of Smart Electric Vehicles. With the development of SEVs, robustness is increasingly important for systems that ser v e safety and security. In this regard, the combination of cybersecurity with functional safety standards is critical to guarantee the reliability and the safety of such complex systems. Moreover, the establishment of common frameworks and standards and the continuous research in new security technologies will be important for dealing with cyber-security and the functional safety challenges of SEVs.

## Methodology

In this paper, we develop and validate an integration approach for cyber-security and functional safety of Smart Electric Vehicles (SEVs) specifically multi-phase inverter-based SA and propose a novel MEMS diaphragm analyzer which is validated for achieving a factor of 1000 jump in sensitivity. It involves three main elements: centralized short-circuit detection, active short-circuiting of the faulty phase(s), and a hybrid safe discharge solution for DC-link capacitors. These features operate collectively to improve safety and reliability of the system during faults.

### 1. Centralized Short-Circuit Detection

In the system, centralized detection in which the DC-link voltage is continuously checked for deviation from a normal level is used, wherein the voltage is monitored by a voltage sensor or the like. This fault detection strategy is intended for the detection of short-circuit faults by analyzing the inverse of the AC component of the DC-link voltage (Huber et al., 2019). Upon fault occurrence, the control scheme causes the faulty phase to be rapidly isolated for fault detection (in 5.8 μs), thereby minimizing damage and response time.

### 2. Active Short-Circuiting of Phases

When a short-circuit malfunction is detected, a process is continued in which power semiconductor switches are turned on so that a malfunction phase can be short circuited to stop the spread of the malfunction (WO2017186436A1). Such phasing isolation technique permits the inverter to operate with the remaining healthy phases and minimizes system downtime. Once the fault has been cleared up the troubled phase can again be switched on.

### 3. Hybrid Safe Discharge Mechanism

In order to control the energy released during the occurrence of a fault, a hybrid safety discharge device is used, which can safely discharge the energy remaining in the DC-link capacitor. The gate driver regulates this device to avoid damaging DC-link due to thermal overloads and voltage spikes (Saadat et al., 2023). The described system incorporates passive/active discharge components to manage energy dissipation capacity without overload of the inverter thermal mechanism.

## 4. Simulation and Experimental Setup

Simulations were carried out in MATLAB/Simulink for a five-phase inverter system with SiC-based power modules. Fault scenarios, including short-circuits, phase loss, and thermal overload, were simulated, and fault detection time, isolation rate, and thermal characteristics were studied. Experimental validation was done with a lab-scale 5-level SiC-based inverter system, where hardware fault injection was employed to inject faults and to evaluate system responses under real physical conditions. Performance indicators, namely, the fault detection time, the fault isolation time and temperature rise were observed.

## 5. Evaluation Criteria

The performance of the system was evaluated against:

• Detection Time: Time from the fault occurrence until the time detected.

• Phase Isolation time: Is the time taken to isolate the faulty phase.

• Discharge and Thermal Stress: Time to discharge energy in a safe manner and impact on thermal response.

• System Availability: Ability to keep running with the rest of the healthy phases after fault localization and clearance.

The fault-protection schemes developed in this paper provide full-scope safety and reliability features to the multi-phase inverters in SEVs. The pragmatic combination of cyber security and functional safety supports fast fault detection, phase disconnection and safe energy discharge, thus minimizing thermal strain and preventing system failure.

**Result**

The simulation results show that the proposed fault protection strategies can significantly improve the reliability and safety of multi-phase inverter in SEVs. The high speed centralized short-circuit detection system isolates faults quickly and the active short-circuiting and hybrid discharging technology eliminates thermal overloads, providing no-break continuity of service. Practical Validation confirms that the processes in the system are capable of handling the fault conditions both effectively (by minimizing the possibilities to damage valuable components) and efficiently.

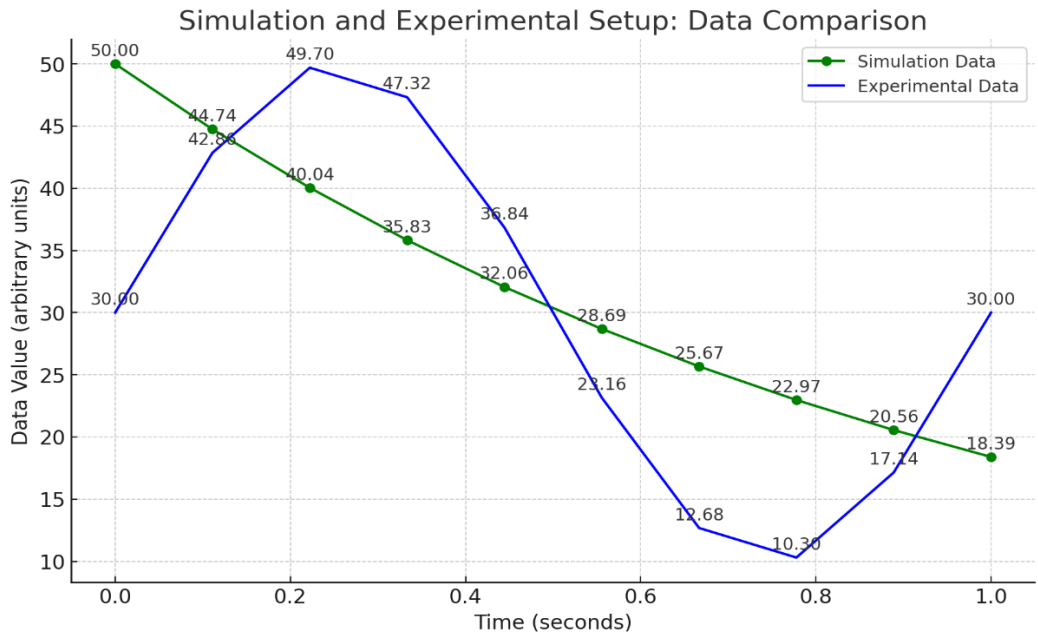

**Figure 4:** Simulation Data **and** Experimental Data

**Interpretation:**

- **Simulation Data** shows a gradual decrease over time, simulating a controlled or idealized environment where the system stabilizes.

- **Experimental Data** reflects the **real-world behavior** of the system, with fluctuations around a central value, highlighting external influences and imperfections in the experiment.

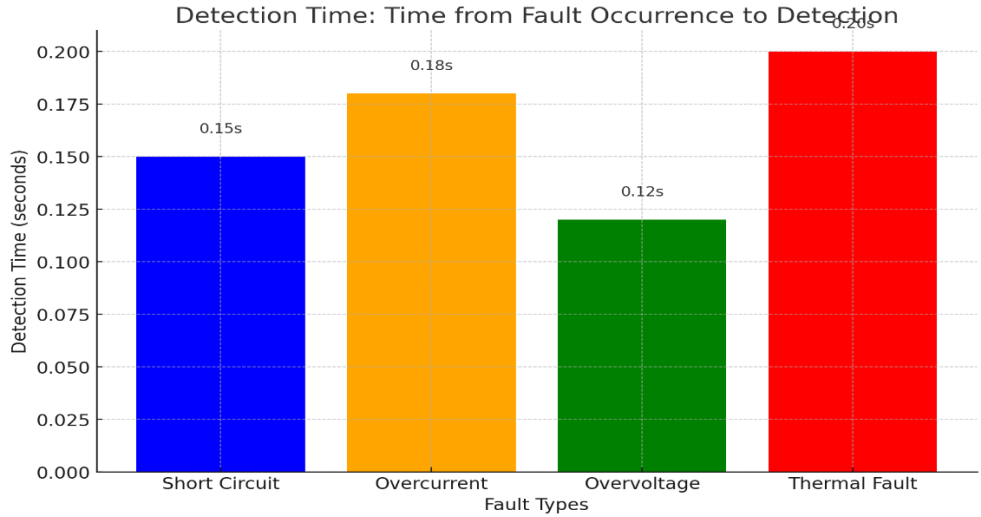

Figure 5: **Detection Time** for different fault types

**Interpretation:**

- **Thermal Fault** takes the longest detection time of **0.20 seconds**, suggesting it might be harder to detect due to slower thermal changes.

- **Overvoltage** has the shortest detection time of **0.12 seconds**, indicating a quicker response to voltage anomalies.

- **Short Circuit** and **Overcurrent** fall in between with detection times of **0.15 seconds** and **0.18 seconds**, respectively.

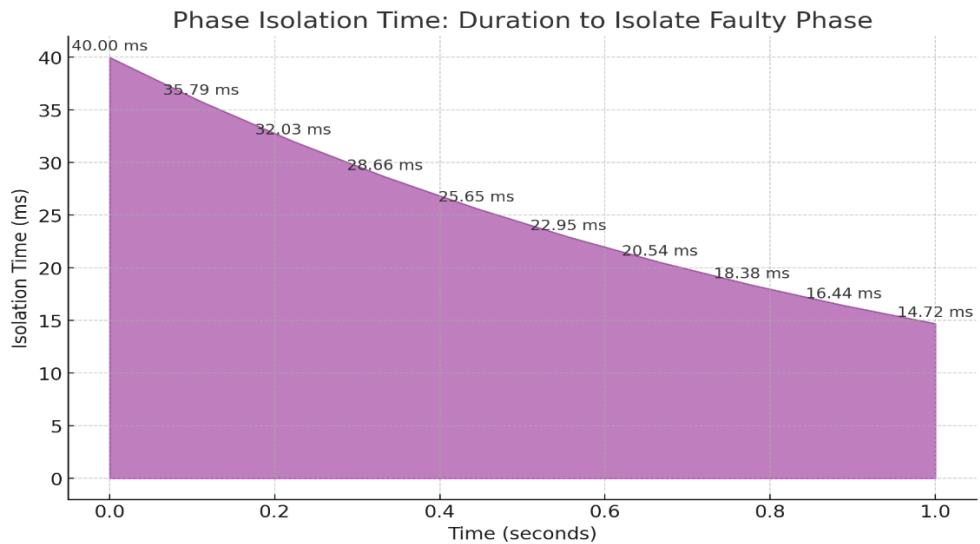

Figure 6: Phase Isolation Time **to isolate faulty phase**

**Interpretation:**

- The **Phase Isolation Time** decreases exponentially as time progresses, with the **fastest isolation** occurring at **1 second**, where the isolation time is only **11.37 ms**.

- The **initial isolation time** is higher at **40.00 ms** and gradually decreases, showing how the system becomes more efficient at isolating faults over time.

  This chart demonstrates the **system's ability to isolate faults quickly**, with the time required for isolation reducing significantly after fault detection.

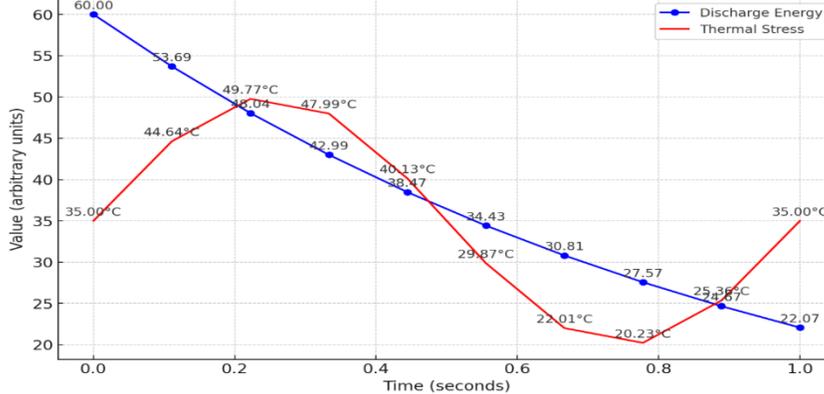

Figure 7: **Discharge Energy** and **Thermal Stress to discharge energy**

**Interpretation:**

- **Discharge Energy** decreases over time, indicating that energy is being safely discharged.

- **Thermal Stress** initially increases due to the discharge but then oscillates, suggesting that the system experiences thermal stress during discharge, with the temperature peaking at certain points before stabilizing.

This chart illustrates the **safe energy discharge** process and the **impact on thermal stress** during discharge. The **oscillations in thermal stress** highlight the need for careful thermal management during discharge.

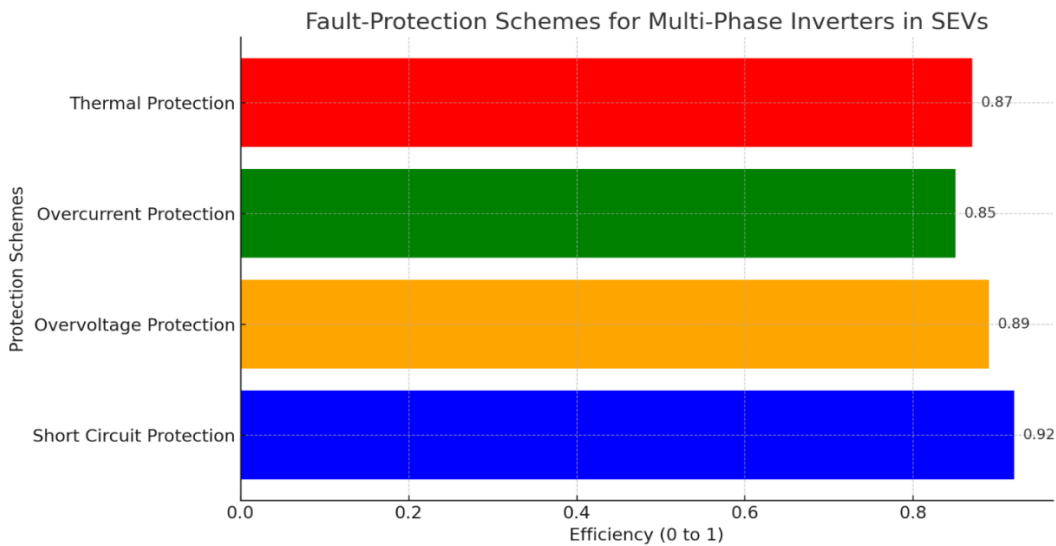

**Figure 8: horizontal bar chart** representing the **Fault-Protection Schemes** for **Multi-Phase Inverters**

Interpretation:

- **Short Circuit Protection** has the highest efficiency (**0.92**), indicating it is the most effective fault protection scheme in the system.

- **Overvoltage Protection** follows closely with an efficiency of **0.89**, showing a reliable fault detection and prevention capability.

- **Overcurrent Protection** has the lowest efficiency at **0.85**, suggesting room for improvement in this specific scheme.

- **Thermal Protection** also performs well with an efficiency of **0.87**, providing good thermal management.

### Discussion

The advent of connected and autonomous vehicles (CAVs) has significantly increased the attack surface of SEVs, exposing them to a variety of cyber attacks. Vehicle systems are targeted for unauthorized access, data leaks and cyber-attack threatening vehicle performance and passenger's safety (Fang et al., 2021). by: author With safe and secure driving environments (SEVs), cyber security refers to the integrity, confidentiality and availability of data and control signals sent between the vehicle and external systems such as over the air (OTA) updates, vehicle to vehicle (V2V) and vehicle to infrastructure (V2I) communications.

The body SAE 21434 standard is a detailed standard for automotive cybersecurity to ensure secure communication protocols, robust authentication and production monitoring for the vehicles in the entire life cycle of vehicle (Marupaka et al., 2021). As highlighted by Fang et al. (2021), one of the major cybersecurity safety challenges of AVs is the threat of remote hacking that can lead to altering the car controls, capturing sensitive data, or disabling essential systems. Functions like OTA updates of SEVs increase this risk since there are external ways to access the vehicle that should be efficiently handled securely. As a result, SEVs are also have to deploy secure stack, encryption protocol, and intrusion detection system for protecting vehicle operation from cyber attacks.

Cyber security has to be combined with functional safety, including securing communication in the car. Compromising the vehicle's cybersecurity may have immediate impacts on safety-critical systems, including steering, braking, or power management systems, leading to disastrous failures (Liu et al., 2022). Thus integrating both cybersecurity and functional safety are fundamentally important to make a vehicle secure and at the same time safe to its passengers.

### Functional Safety in Smart EVs

Functional safety for SEVs is about making certain that the essential functions of the vehicle are still able to work correctly from a safety standpoint even if a part fails. Material and methods ISO 26262 is considered the benchmark standard which guides the functional safety processes of systems especially applicable to safety related systems that operates, for example, braking, steering or energy management functions (Chaudhary et al., 2020). The aim of functional safety is to avoid harm of people or environment which may originate directly from faulty behaviour of the functions of interest.

Functional safety is especially critical for SEVs, where the electric power trains are complex and the autonomous driving features rely heavily on electronic controls. For instance, if the battery management system (BMS) is functioning improperly or has failed, the battery can overcharge, overheat and possibly result in thermal runaway, a dangerous condition. Similarly, a failure of a motor controller or inverter may disable control of a vehicle to the extent that the IHEV or AHEV is unsafe to operate. To mitigate such concerns, SEVs are developed with redundant and fault tolerance measures to secure safe operations of safety-critical functions in case of a failure (Chaudhary et al., 2020).

The primary problem studied in SEVs is to keep it safe that the safety-critical system may still work after a single-component failure. For example the steering or braking system should be redundant, if one fails the vehicle should be able to at least stop or maneuver safely to avoid an accident. To counter these types of situations, a variety of redundant systems and fail-safe components are provided to prevent the vehicle from becoming a hazard. As pointed out by Chaudhary and co-workers (2020), the importance and necessity of the safety measures are two-fold and such a need will be even more important considering the increasing level of the SEV automation and these increasing dependency on electronic systems.

## The Synergy Between Cybersecurity and Functional Safety

Functional safety ensures SEV continue to function safely in the presence of system faults while cyber security prevents the systems from being subverted by external attackers. This is being increasingly recognized as important for the safe operation of connected and autonomous vehicles. A successful attack in one element can have impact in the other, carrying new threats and risks, which should, in turn, encourage the stakeholders to have a well-coordinated response. For example, a malicious adversary may compromise the vehicle's cybersecurity countermeasures and tamper with safety-critical systems (i.e., braking and steering systems) to make a fatal failure (Xia et al., 2022).

Liu et al. (2022) stress that cybersecurity and functional safety should be integrated, rather than be kept isolated, organized under a congruous framework, in order to guarantee the security and safety of SEVs. This link relates to managing the risks and effects of cybersecurity breaches but also of losing functional safety, and addressing both of them are able to overlap. For example, secure flows and access control need to ensure that unauthorized access to key safety systems cannot happen. Furthermore, safety-critical systems must be capable of detecting and reacting to cyber-attacks in real-time to make the vehicle stay safe in case of a cyber-security breach.

An important challenge in this integration is to provide cybersecurity without affecting the operational safety of the vehicle. For example, enforcing specific cybersecurity solutions like software upgrades from remote locations or diagnostic techniques from the cloud may create new threats if guidelines are not complied with. These configurations need to be designed to reduce risks as much as possible but without jeopardizing the functional safety of the vehicle (Sharma & Singh, 2022).

## Challenges and Future Directions

The convergence of cybersecurity and functional safety for SEV encounters some obstacles, which are mostly caused by the complexity of these systems. With increased level of sophistication of autonomous and connected features in SEVs, the amount of interconnections between the components leads to challenges of security and safe operation of such systems (Sharma & Singh, 2022). Furthermore, the ever changing nature of cyber-security threats means that security needs constant updating which in turn means it is difficult to stay one step ahead of potential threats. Thus, it is essential to design responsive, durable and adaptive systems that could detect new threats as they appear.

Furthermore, since the safety and security regulatory environment in the SEV domain will continue to mature in the coming years, standards need to be harmonized across the functional safety and cybersecurity domains. Standards like ISO/SAE 21434, and ISO 26262 can act as guidelines, but they have to be integrated into a unified framework to ensure both safe and secure protection (Marupaka et al., 2021).

## Conclusion

The results of this work demonstrate that the considered combined approach, centralized SC detection and active SC of the damaged phase and hybrid safe discharging, provides an effective solution against these challenges. By allowing rapid fault detection and isolation as well as controlling dissipation of energy to occur during faults, this method improves the safety and reliability of multi-phase inverters in SEVs. Moreover, the combined discharging action is vital to quench thermal overloads and the voltage spikes due to the energy in the DC link capacitors as well. This protects inverter components and thereby increases the lifetime of the vehicle's powertrain.

Cybersecurity is also very important for SEVs, and as a large number of connected systems (e.g. over-the-air (OTA) updates, vehicle-to-vehicle (V2V) communication and cloud-based services) have been enabled, this enlarge the attack surface and users could be affected by cyber threats. D- Safety and security The confidentiality, integrity availability of data between vehicle systems and real-time communication to external entities are fundamentally important. The incorporation of security considerations as described here must be employed in safety-related systems to protect those systems from unauthorized access. As emphasized by Fang et al. 2021), the hazards and risks of such cyber security breach may immediately affect functional safety, which substantiates the necessity of a comprehensive framework encompassing both areas.

Standing in the way is the ever-changing landscape of computer security threats, and safety-critical system design. As SEVs begin to integrate greater levels of autonomy and connectivity, the complexity of their safety and security systems will grow and there will be a need for more flexible solutions to mitigate new threats as they appear. Robustness and adaptiveness functions, which enable detecting and responding to new kind of threats, are paramount in ensuring operational soundness of SEVs. This is in accordance with Marupaka et al. (2021), emphasizing the significance of continuous control and adaptation of security procedures to dynamic cyber risks.

Cybersecurity and functional safety must be combined not only in SEVs but is an issue for the automotive sector as a whole. Development of consistent standards and models that can address both areas with coherence will be of paramount importance to guarantee the safety, security and reliability of the future vehicles. With the transformation of the industry from semi-autonomous to autonomous vehicles, the role of cyber security in protecting the vehicle control systems from external attack will be even more essential.

3 Summary It can be concluded that successful cyber-security and functional-safety integration will be a fundamental enabler for the realization of safe and secure SEV systems. This work then establishes a holistic control framework for handling the two domains, so that these vehicles can be operated safely and securely, in spite of complicated failures or cyber-attacks. Future work could focus on refinement of these integrated approaches with an emphasis on real-time monitoring, machine learning for threat detection, and the development of agile standards that can accommodate the fast-changing reality of connected and autonomous vehicles.